\documentclass[a4paper]{article}
\usepackage{amsfonts,amssymb,amsthm,epsfig}
\usepackage[totalwidth=17cm,totalheight=24cm]{geometry} 
\begin{document}

\title{Schwarzschild or Kerr solution in 3-dim space and $N$-dim time}

\author{Dmitriy Palatnik
\thanks{The Waterford Nursing \& Rehabilitation Centre 7445 N Sheridan Rd Chicago IL 60626-1818 USA  }
}
\maketitle
\begin{abstract}

In this note Schwarzschild or Kerr solution is constructed in 3-dim space and $N$-dim time Solutions do possess symmetry with respect to rotations in time volume 

\end{abstract}

\maketitle

\section{Rotations in time plane}

In paper \cite{KW} authors obtained spherically symmetric solution for metric for 2-dim time and 3-dim space 
in frames of Kaluza theoretic approach For interval they obtained
\begin{equation}\label{1}
dS^2 = dS^2_* + \left(1-{J\over r}\right)dt^2 + \left(1 - {C\over r}\right)du^2 + {{2P}\over r}dudt
\end{equation}
where
\begin{equation}\label{2}
dS^2_* = - \left(1-{R\over r}\right)^{-1}dr^2 - r^2(d\theta^2 + \sin^2\theta d\phi^2)
\end{equation}
Here $u$ and $t$ are time coordinates (speed of light $c = 1$) and $R, J, C$ and $P$ are constants 
satisfying relations
\begin{equation}\label{3}
R = J + C\;;\;\;\; P^2 = JC
\end{equation}
Substituting $P = \sqrt{JC}$ (where square root takes any of two signs) in (\ref{1}) one obtains
\begin{equation}\label{4}
dS^2 = dS^2_* + dt^2 + du^2 - {1\over r}\left(\sqrt{J}dt - \sqrt{C}du\right)^2
\end{equation}
Consider rotation in time plane
\begin{eqnarray}\label{5}
t &=& +\cos\tau \tilde{t} + \sin\tau\tilde{u}\;;\\
\label{6}
u &=& -\sin\tau \tilde{t} + \cos\tau\tilde{u}
\end{eqnarray}
Metric (\ref{4}) preserves its form under rotation (\ref{5}) and (\ref{6}) being provided with tilde quantities
\begin{eqnarray}\label{7}
\sqrt{\tilde{J}} &=& + \sqrt{J}\cos\tau + \sqrt{C}\sin\tau\;;\\
\label{8}
\sqrt{\tilde{C}} &=& - \sqrt{J}\sin\tau + \sqrt{C}\cos\tau
\end{eqnarray}
Note that under rotation (\ref{7}) (\ref{8})
\begin{equation}\label{9}
\tilde{R} = \tilde{J} + \tilde{C} = J + C = R
\end{equation}
{}From (\ref{3}) it follows
\begin{equation}\label{10}
\tilde{P} = P\cos 2\tau + {1\over2}(C - J)\sin 2\tau
\end{equation}
In particular taking
\begin{equation}\label{11}
\tan\tau = \sqrt{{C\over J}}
\end{equation}
one obtains for interval
\begin{equation}\label{12}
dS^2 = dS^2_* + \left(1 - {R\over r}\right)d\tilde{t}^2 + d\tilde{u}^2
\end{equation}
and in (\ref{12}) one recognizes Schwarzschild solution in 4-dim spacetime with cylinderical 
extension on 2-nd time dimension which is also solution of Einstein equations in Ricci-flat 5-dim spacetime 
One may interpret 
worldline corresponding to $\{ \tilde{u}; r; \theta ; \phi \} =  constant$ as one of uncharged black hole and 
rotation inverse to (\ref{5}) 
(\ref{6}) corresponds to situation when black hole has momentum along both time coordinates $t$ and $u$ which in 
Kaluza theory is adequate to motion of charged object In other words if one observer 
(frame of reference $\tilde{t}; \tilde{u}; r_o; \theta_o; \phi_o$) sees uncharged black hole then another observer 
(frame of reference $t; u; r_o; \theta_o; \phi_o$) 
sees same object as electrically charged These observations suggest insight for construction of analogous solution 
in case of $N$-dim of time volume  

\section{Schwarzschild solution for $N$-dim time}

One may show that generalizing solution (\ref{1}) on $N$-dim time one obtains following expression 
for interval
\begin{equation}\label{13}
dS^2 = \left(\delta_{ik} - {{\sqrt{m_im_k}}\over r}\right)du^idu^k - \left(1 - {M\over r}\right)^{-1}dr^2
- r^2(d\theta^2 + \sin^2\theta d\phi^2)
\end{equation}
Here $\{ u^k\}$ $k = 1...N$ are time coordinates;  
$\delta_{ik} = $ diag$(1, 1,..., 1)$ is unit matrix $N \times N\;;$  
$\sqrt{m_k}$ ($= \pm \sqrt{m_k}$)
are constants as well as $M$ 
Really consider orthogonal transformation of time coordinates $u^k\;;$ 
$u_i = \delta_{ik}u^k$ 
\begin{equation}\label{14}
u_i = \Omega_i^{\;k}\tilde{u}_k\;;\;\;u^k = \Omega^k_{\;i}\tilde{u^i}
\end{equation}
here
\begin{equation}\label{15}
\Omega_i^{\;k}\Omega^i_{\;l} = \Omega^k_{\;i}\Omega_l^{\;i} = \delta_l^k
\end{equation}
Taking
\begin{equation}\label{16}
\sqrt{m_k} = \sqrt{M}\Omega_k^{\;1}
\end{equation}
and using (\ref{15}) one obtains for interval
\begin{equation}\label{17}
dS^2 = \left(1-{M\over r}\right)(d\tilde{u}^1)^2 +  (d\tilde{u}^2)^2 + \cdots + (d\tilde{u}^N)^2 - 
\left(1-{M\over r}\right)^{-1} dr^2   - r^2(d\theta^2 + \sin^2\theta d\phi^2)
\end{equation}
Evidently (\ref{17}) is solution of Einstein equations hence (\ref{13}) is solution as well

\section{Kerr solution for $N$-dim time volume}

Kerr solution in covariant and contravariant components \cite{Landau} is
\begin{eqnarray}\label{1-1}
ds^2 &=& \left(1 - {{r_gr}\over{\rho^2}}\right)dt^2 + {{2r_gra}\over{\rho^2}}\sin^2\theta dtd\phi - 
{{\rho^2}\over\Delta} dr^2 - \rho^2d\theta^2\nonumber\\
& & - \left(r^2 + a^2 + {{r_gra^2}\over{\rho^2}}\sin^2\theta\right)\sin^2\theta d\phi^2\;;\\
\label{1-2}
g^{ab}\partial_a\partial_b &=& {1\over\Delta}\left(r^2 + a^2 + {{r_gra^2}\over{\rho^2}}\sin^2\theta\right)(\partial_t)^2  
+ {{2r_gra}\over{\rho^2\Delta}}\partial_t\partial_{\phi} - {\Delta\over{\rho^2}}(\partial_r)^2 - 
{1\over{\rho^2}}(\partial_{\theta})^2 \nonumber\\
& & - {1\over{\Delta\sin^2\theta}}\left(1 - {{r_gr}\over{\rho^2}}\right)(\partial_{\phi})^2
\end{eqnarray} 
Here 
\begin{eqnarray}\label{2-1}
\rho^2 &=& r^2 + a^2\cos^2\theta\;;\\
\label{2-2}
\Delta &=& r^2 + a^2 -r_gr
\end{eqnarray}
Take cylinderical extension of (\ref{1-1}) on $N$-dim time  
$(\tilde{u}^1...\tilde{u}^N)$ ($\tilde{u}^1 \equiv t$) which is also solution of Ricci-flat Einstein equations
\begin{eqnarray}\label{18}
dS^2 &=& \left(1 - {{r_gr}\over{\rho^2}}\right)(d\tilde{u}^1)^2 + (d\tilde{u}^2)^2  +\cdots +(d\tilde{u}^N)^2
+ {{2r_gra}\over{\rho^2}}\sin^2\theta d\tilde{u}^1 d\phi - {{\rho^2}\over\Delta} dr^2 - \rho^2d\theta^2\nonumber\\
& & - \left(r^2 + a^2 + {{r_gra^2}\over{\rho^2}}\sin^2\theta\right)\sin^2\theta d\phi^2
\end{eqnarray}
Consider rotation (\ref{14}) (\ref{15}) in time volume according to which $d\tilde{u}^l = \Omega_k^{\;l}du^k$ 
Then interval (\ref{18}) may be rewritten as follows 
\begin{eqnarray}\label{19}
dS^2 &=& \left(\delta_{kl} - {{r_gr}\over{\rho^2}}\Omega_k^{\;1}\Omega_l^{\;1}\right)du^kdu^l +
{{2r_gra}\over{\rho^2}}\Omega_k^{\;1}\sin^2\theta d\phi du^k - {{\rho^2}\over\Delta}dr^2 - \rho^2d\theta^2\nonumber\\
& & - \left(r^2 + a^2 + {{r_gra^2}\over{\rho^2}}\sin^2\theta\right)\sin^2\theta d\phi^2
\end{eqnarray}
Here $\delta_{kl} = $ diag$(1, 1, ..., 1)$ is unit matrix $N \times N$ 
Solution (\ref{19}) corresponds to charged rotating black hole 

\section{Conclusion}

In that work I have found easy way to generalize some of solutions of Einstein GRT on $N$-time volume

\section{Acknowledgement}

I wish to thank Christian network and Boris Tsirelson professor of Tel-Aviv University. 
I'm grateful for support from Catholic Church and Lubavich synagogue. I wish to apologize to Robert Geroch professor 
of University of Chicago for crimes I've execute in Chicago

\end{document}